# An agile laser with ultra-low frequency noise and high sweep linearity

**Haifeng Jiang[1,*], Fabien Kéfélian[2], Pierre Lemonde[1], André Clairon[1] and Giorgio Santarelli[1]**

[1]*Laboratoire National de Métrologie et d'Essais–Système de Références Temps-Espace, Observatoire de Paris, UPMC and CNRS, 61 Avenue de l'Observatoire, 75014 Paris, France*
[2]*Laboratoire de Physique des Lasers, Université Paris 13 and CNRS, 99 Avenue Jean-Baptiste Clément, 93430 Villetaneuse, France*

[*]*jiang.haifeng@obspm.fr*

**Abstract:** We report on a fiber-stabilized agile laser with ultra-low frequency noise. The frequency noise power spectral density is comparable to that of an ultra-stable cavity stabilized laser at Fourier frequencies higher than 30 Hz. When it is chirped at a constant rate of ~ 40 MHz/s, the max non-linearity frequency error is about 50 Hz peak-to-peak over more than 600 MHz tuning range. The Rayleigh backscattering is found to be a significant frequency noise source dependent on fiber length, chirping rate and the power imbalance of the interferometer arms. We analyze this effect both theoretically and experimentally and put forward techniques to reduce this noise contribution.

# 1. Introduction

Simultaneous achievement of low frequency noise operation and precise, fast and linear tunability is a challenge for laser technology. These features are key requirements for many applications: coherent light detection and ranging (lidar) [1], high resolution spectroscopy [2], optical tracking oscillator (including phase coherent tracking of optical signal from a satellite [3]), heterodyne high resolution optical spectrum analyzer, optical processing of radio-frequency signals [4], coherent manipulation of atoms for quantum information storage [5], low noise interferometric sensors [6] or optical frequency-modulated continuous-wave reflectometry [7].

Lasers with sub-hertz line-width and fractional frequency instability around $10^{-15}$ for 0.1 s to 10 s averaging time are currently realized by locking onto an ultra-stable Fabry-Perot cavity using the Pound-Drever-Hall method [8-11]. However, this method requires fine alignment of free space optical components, tight polarization adjustment and spatial mode matching. Moreover, ultra-stable cavities are relatively expensive, bulky and fragile. Finally, this method is not convenient to realize large range sweeping.

In the last two decades, several research groups have developed another approach for short term laser frequency noise reduction, using large arm length unbalance fiber interferometers [12-18]. Recently, we have dramatically improved the performance of such systems by using a longer fiber spool and a heterodyne detection technique [19]. It was also shown that this approach allows well controlled laser frequency tuning without acting on the interferometer reference [13,16,17,20].

In this paper, we demonstrate the ultra-low frequency noise performance of a chirped laser stabilized to a fiber-based interferometer. We analyze the noise sources and fundamental limits of the method, in particular the impact of the Rayleigh backscattering (RBS), which constitutes the present limit of our system.

## 2. Operation principle and experimental setup of the agile laser

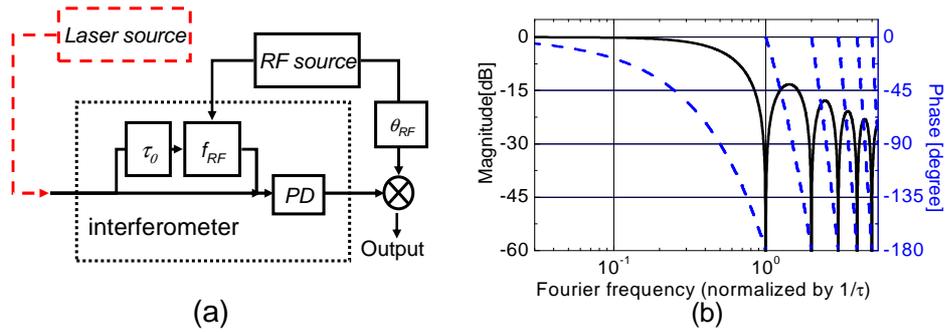

Fig.1. (a) Heterodyne interferometer, PD (Photodiode); (b) transfer function of the interferometer: relative magnitude $|T(f)|/2\pi\tau_0$ [black solid line] and phase [blue dashed line] of frequency response of the interferometer.

The frequency of a laser can easily be locked onto the optical length $L_0$ of a fiber by inserting the latter into one arm of a Mach-Zehnder or Michelson interferometer which can then be used as a frequency discriminator.

However, in this homodyne configuration the locked laser frequency $\nu_0$ is limited to a discrete set of values $(1/2+k) c/2L_0$, where $k$ is an integer and $c$ the speed of light in vacuum, corresponding to the quadrature condition of the arm outputs $\cos(2\pi\nu_0 L_0/c)$ to be 0. The locked laser frequency can be tuned within a small range by adding a variable offset to the error signal, but large range continuous tuning requires a corresponding tuning of $L_0$.

Insertion of a frequency shifter in one arm of the interferometer as shown in Fig.1a enables heterodyne detection. One of many advantages of this configuration is to allow the tuning of $\nu_0$ without modifying the fiber length according to the locking condition:

$$\cos[2\pi\nu_0\tau_0 - \theta_{RF}(t)] = 0, \qquad (1)$$

where $\tau_0 = L_0/c$ is the unbalance time delay of the interferometer and $\theta_{RF}(t)$ the phase difference between the modulation and demodulation signals. The frequency equals to the sum of $(1/2+k)/2\tau_0$ and $\theta_{RF}(t)/2\pi\tau_0$, can then be tuned by controlling $\theta_{RF}(t)$, which can be done more precisely and with a greater dynamic range than a control of $L_0$.

The transfer function relating the laser frequency fluctuations $\delta\nu$ to the heterodyne signal phase fluctuations is

$$T(f) = \frac{(1-e^{-i2\pi\tau_0 f})}{if} \text{ rad/Hz,} \qquad (2)$$

where $f$ is the Fourier frequency. This function is plotted both for phase and magnitude in Fig. 1b. At low frequencies ($f \ll 1/\tau_0$), the magnitude is approximately equal to $2\pi\tau_0$, which means that a longer fiber increases the discriminator slope. Standard deviation of the total fiber length fluctuations $\delta L$, arising from spatially uncorrelated local length fluctuations, is proportional to $L_0^{1/2}$, while the error signal scales as $L_0$. Consequently, a longer fiber makes the system less sensitive to the distributed fiber noise. However, for integer values of $1/\tau_0$, the interferometer transfer function has null amplitude corresponding to zero gain points in the loop response. Maintaining a bandwidth larger than $1/\tau_0$ is feasible, but requires a complex servo-loop design [21]. If we restrict the bandwidth to below $1/\tau_0$, there is a trade-off between the lower noise floor for low Fourier frequencies ($\ll 1/\tau_0$) and the larger bandwidth of the frequency noise rejection.

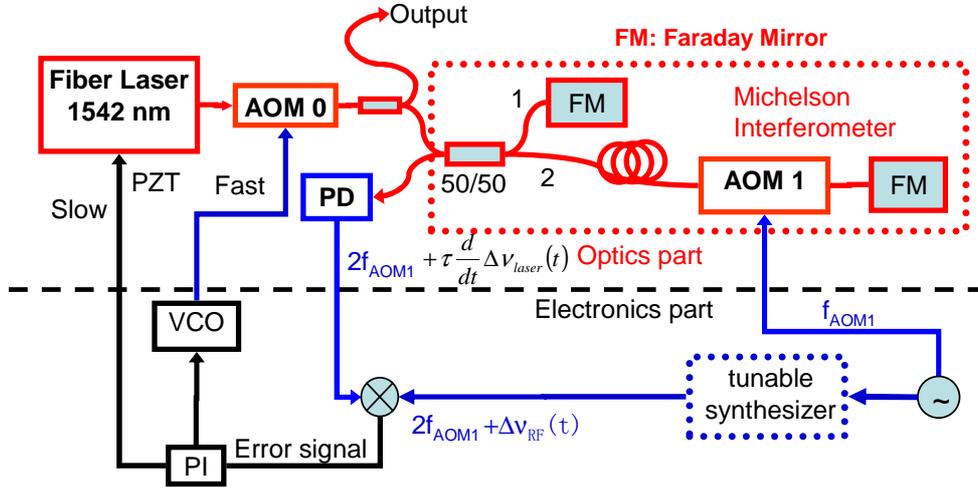

Fig. 2. Experimental setup frame scheme: VCO (Voltage Controlled Oscillator), AOM (Acousto-Optical Modulator), PD (photodiode), PI (proportional-integral control system), PZT (piezo-electric transducer).

Fig. 2 shows the experimental setup, where the optical signals, radio-frequency (RF) signals and low frequency signals are drawn in red, blue and black respectively. The main difference from our previous setup [19] is the use of a low phase noise tunable synthesizer instead of a RF frequency multiplier. An acousto-optic modulator (AOM1) is placed into one arm of the interferometer leading to a heterodyne beat-note signal at frequency $2f_{AOM1}$ at the output of the photodiode. By using a Michelson interferometer configuration with two Faraday mirrors, the polarization of the output waves in the output port is automatically aligned, leading to a maximum beat-note signal without any polarization adjustment. The tunable synthesizer provides an RF signal with frequency $2f_{AOM1}+\Delta\nu_{RF}(t)$. This signal is used to demodulate the heterodyne signal. According to (1), when the control loop is closed the frequency offset $\Delta\nu_{RF}(t)$ will then induce a laser frequency change $\Delta\nu_{laser}(t)$ given by

$$\Delta\nu_{laser}(t) = \frac{\Delta\theta_{RF}(t)}{2\pi\tau_0} = \frac{\int_0^t \Delta\nu_{RF}(t')dt'}{\tau_0}, \qquad (3)$$

This expression shows that a constant frequency offset $\Delta\nu_{RF}$ generates a linear laser frequency sweeping with chirp rate $\Delta\nu_{RF}/\tau_0$.

The fiber delay line is a 2.5-km SMF-28 fiber spool ($\tau_0 \sim 25$ µs), placed in an air-sealed can, to reduce the thermal and mechanical noises coupled from the environment. The interferometer is housed in an aluminum box with thermal isolation (Mylar and thermoplastic polymer foils), whose size is 260 mm x 260 mm x 165 mm, sitting on a commercial compact passive vibration isolation platform. The servo-loop has a bandwidth of about 20 kHz with 3 integral correction stages to reduce the error during frequency sweep. Control of the laser frequency is realized by using the piezoelectric transducer (PZT) stretcher port of the commercial fiber laser for slow corrections on large range, and AOM0 for fast corrections. In this experiment, all interferometer components are pigtailed off-the-shelf, which makes the system alignment-free, simple and robust.

### 3. Measurements and discussion

*3.1 Frequency noise and instability of the non-chirped laser*

In non-chirped operation, the RF beat-note signal between the fiber-stabilized laser and a high-finesse Fabry-Perot cavity-stabilized laser described in [22] is frequency-to-voltage converted and analyzed by a fast Fourier transform analyzer. Frequency noise measurements are presented in Fig. 3.

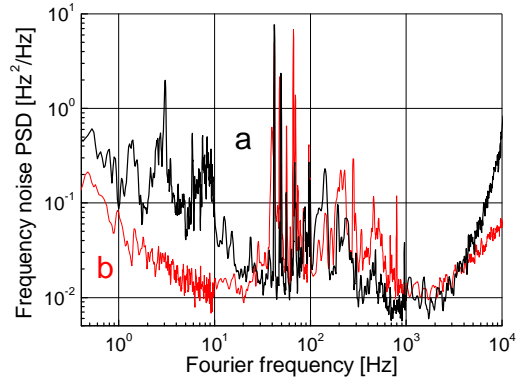

Fig. 3. Frequency noise PSD of lasers beat-note between (a) fiber-stabilized laser and an ULE-cavity stabilized laser, (b) two ULE-cavity stabilized lasers. The frequency noise PSD of the free running laser is given in [21]. Note that no drift is removed in these measurements.

For Fourier frequencies between 30 Hz and 3 kHz, trace (a) coincides with trace (b), which means that, in this frequency range, the frequency noise of the fiber-stabilized laser is at least as low as the frequency noise of one of the ULE-cavity stabilized lasers. At higher frequencies, the noise increases due to the limited gain of the servo-loop. At low frequencies, the fiber-stabilized laser has more noise due to the temperature fluctuations and low frequency mechanical vibrations. The thermal sensitivity of the interferometer is $\sim 10^{-5}$ /°C and we have measured the relative vibration sensitivity to be a few $10^{-10}$/m/s$^{-2}$. Compared to our previous results obtained with a 1-km fiber [19], the performances in the decade 1 Hz-10 Hz have been improved by up to 10 dB. This improvement is mainly due to the use of the air-sealed can. The fiber fundamental thermal floor calculated using Wanser's approach [23] is about $8 \times 10^{-4}$ Hz$^2$/Hz still 10 dB below our measured frequency noise. Using the beat-note signal frequency recorded by a counter, we evaluate the laser frequency stability using the square root of the Allan variance [24,25]. After removing a linear drift of about 1 kHz/s for a measurement of 5 minutes duration, the Allan deviation is close to $10^{-14}$ for integration times ranging from 0.1 s to 1 s.

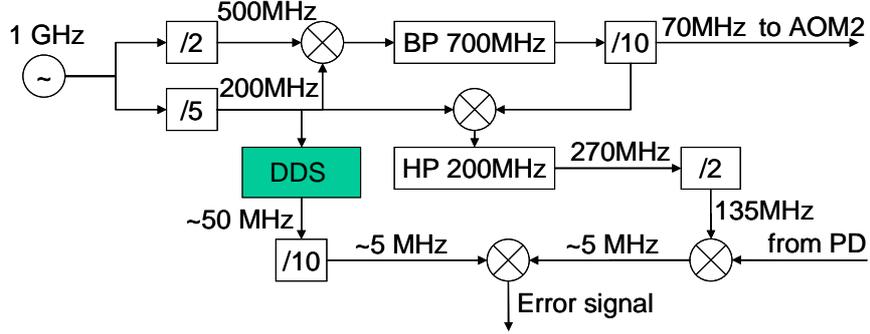

Fig. 4. The scheme of the tunable synthesizer, BP (band-pass filter), HP (high-pass filter), DDS (direct digital synthesizer).

*3.2 Tunable synthesizer design*

A critical part of the electronic system is the tunable synthesizer, which is specially designed for this experiment. Indeed, the phase noise introduced by the tunable synthesizer, $\delta\theta_{RF}$, induces an optical frequency noise power spectral density $S_{\delta\theta\_RF}(f)/(2\pi\tau_0)^2$. The scheme of this synthesizer is shown in Fig. 4. All the signals are synthesized from a common 1 GHz RF source. The AOM1 frequency is 70 MHz, thus the heterodyne beat-note signal is around 140 MHz. The tuning function is realized by using a Direct Digital Synthesizer (DDS) (AD9852) with a 200 MHz clock frequency reference. To obtain a low phase noise tunable local signal a double conversion technique is used. We first down-convert the 140 MHz heterodyne signal to an intermediate frequency of ~5 MHz and then we mix it with a tunable DDS output signal at 5 MHz+$\Delta\nu_{RF}(t)$ to produce a base-band error signal. With this technique, the electronic phase noise is about -106 dBrad$^2$/Hz at 1 Hz offset frequency, which is converted to a laser frequency noise of $10^{-3}$ Hz$^2$/Hz (Eq. 3). For frequencies higher than 10-Hz. the phase noise is well below -110 dBrad$^2$/Hz. Hence, the tunable synthesizer phase noise will not degrade the performance of the laser.

*3.3 Frequency noise of the chirped laser*

The frequency noise of the chirped laser cannot be simply measured with the same frequency-to-voltage technique used for the non-chirped laser. Indeed the operating range of the frequency-to-voltage converter is limited to about 1 MHz much less than the few hundreds MHz sweeping range in our experiment. Extending the range by digital frequency division would severely degrade the noise measurement floor. Consequently, we have used two alternative solutions to overcome the limitation of the frequency to voltage technique. One is based on the sampling of the frequency by a high resolution counter, which will be shown later to be interesting for low Fourier frequencies (<30 Hz). But first we exploit a second 1-km Michelson frequency-shifted interferometer to convert the laser frequency noise into RF phase noise with a larger bandwidth. With this technique, a linear optical frequency sweep is converted into a constant RF frequency shift. The measurement scheme is shown in Fig. 5a. The phase noise of the beat-note signal at the output of the measurement interferometer is evaluated with a standard phase noise measurement technique. We use a low phase noise quartz oscillator weakly phase locked on the RF output signal of the measurement interferometer (~1 Hz bandwidth). Thus, we obtain the phase noise at the output of the phase detector for Fourier frequencies larger than the control bandwidth which is measured using a FFT analyzer. The phase noise PSD is then converted into optical frequency noise PSD using the scaling factor $1/(2\pi\tau_l)^2$, where $\tau_l$ is the 1-km interferometer delay. With this technique, the noise PSD is the sum of both interferometer contributions and the measurement floor is relatively high, about 2 Hz$^2$/Hz. Nevertheless it was sufficient to reveal the effect of Rayleigh backscattering.

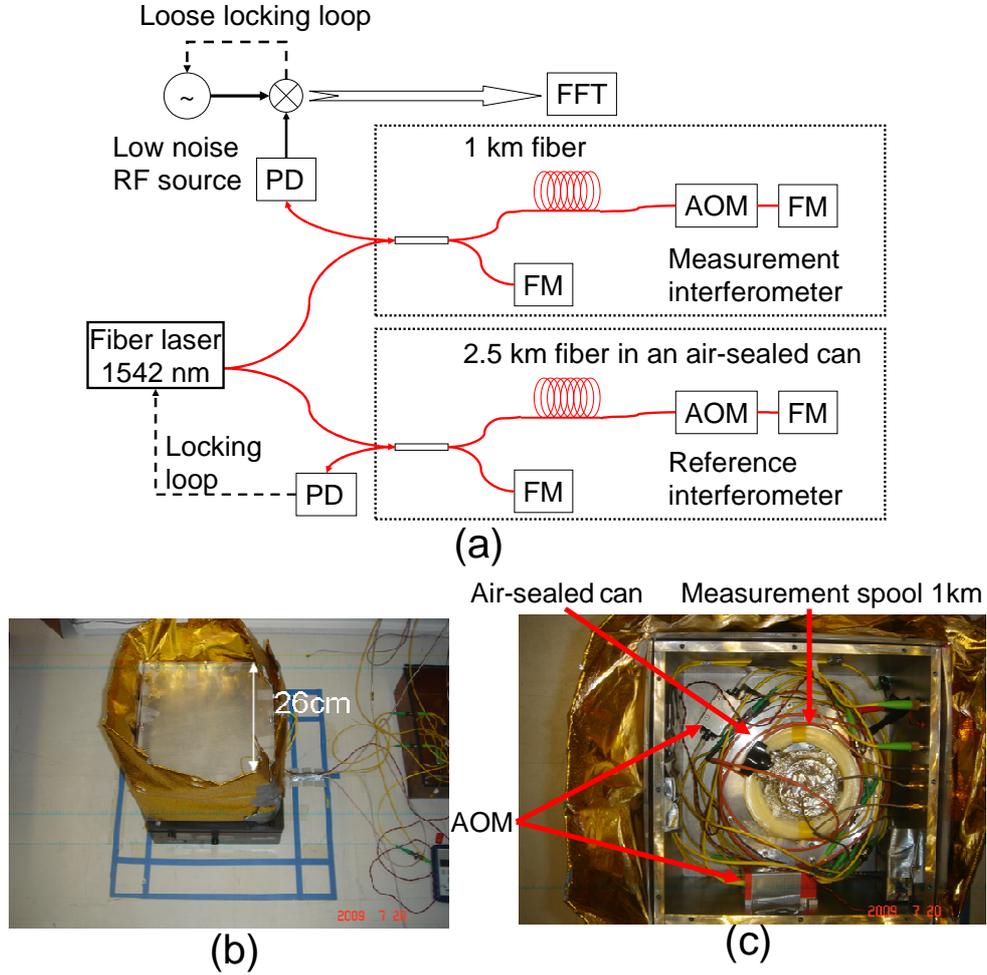

Fig. 5. (a) Scheme of the frequency noise measurement of a chirped laser using a second demodulation interferometer, (b) and (c) views of the Michelson interferometers set-up.

The measured frequency noise PSD is plotted in Fig. 6 for chirping rates 1 MHz/s and 2MHz/s. The spectra are significantly degraded with respect to the non-chirped situation and exhibit a quite peculiar shape characterized by two steps and several bright peaks. We attribute the bright peaks to localized stray reflections in the interferometers. The steps are a clear sign of Rayleigh backscattering in both the reference and measurement fibers. Indeed, as shown in Appendix 1, frequency noise due to RBS in one of the interferometers is expected to be

$$\begin{cases} S_\nu(f) = \left(\frac{a_1}{a_2}\right)^2 \frac{a_L a_B c}{8n} \frac{1}{(2\pi\tau_0)^2} \frac{1}{\nu'}, f \leq \tau_0 \nu' = \Delta\nu_{RF}, \\ S_\nu(f) = 0, f > \tau_0 \nu' \end{cases} \qquad (4)$$

where $a_1/a_2$ is the ratio of the field amplitudes coupled to both arms of the interferometer (coupling ratio), $a_L$ (~3.10$^{-5}$/m) the Rayleigh loss of the fiber per unit of length, $a_B$ (~1/500) the backward coupling coefficient [26], $n$ the refractive index of the fiber, $\nu'$ the laser tuning rate. The level and frequency width of the steps depend on the chirping rate and the delay time of each interferometer. The observed widths of steps match exactly the value expected from Eq. 4 for both chirp rates.

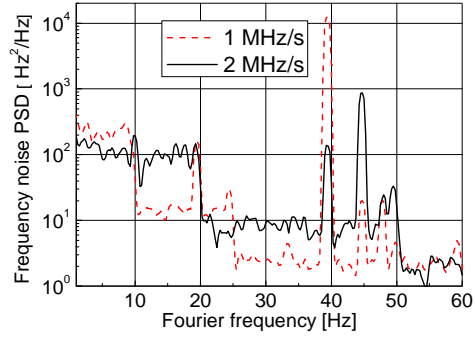

Fig. 6. Frequency noise power density of chirped laser at different chirping rates (a linear frequency drift is removed).

By fitting the measurements presented in Fig. 6 with this expression (Eq. 4), we found $a_1/a_2 = 0.7$ and $a_1/a_2 = 0.5$ for the measurement and reference interferometer respectively. These values are smaller than 1 which can be explained by the additional splicing loss in the long arms and the additional bending loss at the input of the air-sealed can for the reference interferometer.

*3.4 Reduction of the Rayleigh backscattering effect and fundamental limits*

An obvious way of reducing the degradation due to RBS is to minimize $a_1/a_2$ by inserting attenuators at the entrance of the longer arm. However a small gain is expected and would result from a trade-off between the shot-noise and Brillouin scattering which limits the input power of the system. Another possibility would be to use a Mach-Zehnder instead of the present Michelson configuration. However, it would be sensitive to polarization fluctuations hardly compatible with a robust operation, unless adding a polarization control system which increases the experimental complexity.

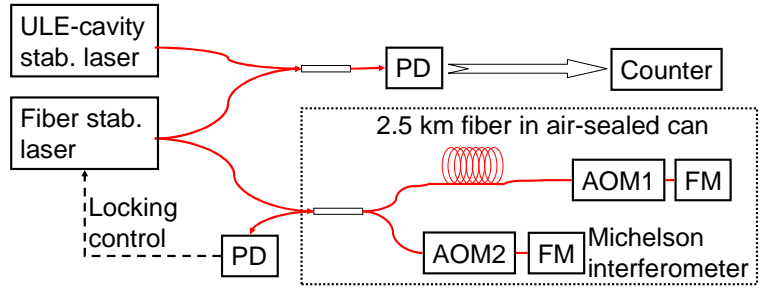

Fig. 7. Enhanced Michelson interferometer scheme with single RBS immunity

A better solution is to add another AOM (AOM2 in Fig 7.) in the short arm of the Michelson interferometer. By detecting the RF signal at twice the difference frequency between both AOMs, one is not sensitive to single RBS from the long arm which is no longer at the same frequency as the short arm signal since AOM1 is at the end of the long arm. The experimental scheme is shown in Fig. 7. To measure the frequency noise of the chirped laser with a 2-AOM configuration, which is expected to be lower than the previous system, we compare the chirped laser signal with the cavity stabilized laser signal. The beat-note signal frequency is sampled by a frequency counter with a gate time of 1 ms, and the frequency noise spectrum is obtained by fast Fourier transform. This frequency noise measurement method exhibits a lower noise floor especially at low Fourier frequencies (<100 Hz) than the previous double interferometer method we used. However, the spectral analysis range is limited to half the inverse of the counter gate time and the spectral power density is degraded by aliasing producing an excess noise on the order of $10^{-4} f^{2}$ Hz$^2$/Hz. Fig. 8 shows the laser frequency noise spectrum measurements. In non-chirped conditions, the measured noise is close to that of Fig. 3 up to about 20 Hz. For higher Fourier frequencies, aliasing starts playing a significant role and limits the measurement noise floor. In chirped conditions, the noise level reaches the measurement floor at Fourier frequencies higher than $\Delta v_{RF}$. At frequency $\Delta v_{RF}$ however each spectrum still exhibits a step, which is characteristic of the effect of single RBS. We attribute this feature to a stray reflection in the long interferometer arm. Indeed, the RF signal at the interferometer output displayed on a RF spectrum analyzer shows a bright line at twice the AOM2 frequency which indicates a stray reflection of optical power of about 1-2% located in the long arm of the interferometer. In the case of stray reflections located between the optical splitter and AOM1, a fraction of the single RBS signal propagates towards the Faraday mirror and experiences twice the AOM1 frequency shift. It consequently contributes to the beat-note signal at frequency $2(f_{AOM1}+f_{AOM2})$ and induces a frequency noise which follows Eq. 4 where $(a_1/a_2)^2$ is replaced by the parasitic reflection coefficient. In the future, the stray reflections can be reduced to a level $\sim 10^{-4}$ or better, which correspond to the level of Rayleigh backscattering in a 2-km fiber.

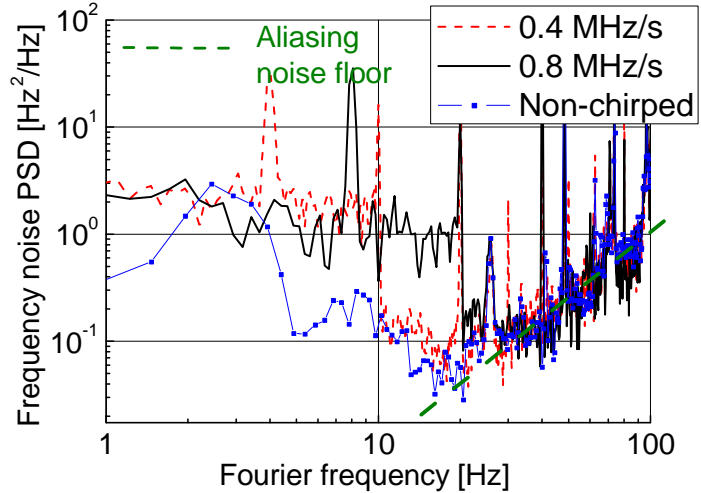

Fig. 8. Frequency noise of single RBS suppressed fiber-stabilized laser for various chirp rates.

Ideally in this 2-AOM configuration, only the second order back-scattering effect (double RBS) leads to a degradation which is about 4 orders of magnitude smaller than in the previous one (see below). Frequency noise induced by double RBS cannot be avoided and therefore constitutes a fundamental frequency noise floor for this technique. The expression of this noise which is derived in appendix 2 is:

$$\begin{cases} S_v(f) = \left(\dfrac{a_L a_B c}{n}\right)^2 \dfrac{1}{32\pi^2 \tau_0 v'} \left(1 - \dfrac{f}{\tau_0 v'}\right), f \leq \tau_0 v' = \Delta v_{RF} \\ S_v(f) = 0, f > \tau_0 v' \end{cases} \quad (5)$$

Noise due to double RBS behaves differently to that from single RBS. The distribution of frequency noise over [0, $\Delta\nu_{RF}$] is no longer uniform but decreases linearly with frequency as shown in Fig. 2.1 (Appendix 2). It is worth noting that the total integrated frequency noise power is independent of the chirp rate and the fiber length for both noise sources. The ratio of the integrated noise power between single RBS and double RBS is $1/(a_L a_B L_f)$ ($6.7 \times 10^3$ for our case).

*3.5 Linearity of the chirped laser*

As an agile laser, the tuning linearity is an important feature [27]. Indeed, a chromatic dispersion $D(\lambda)$ of the fiber leads to a change in delay time. As a result, the chirping rate becomes

$$\nu' = \frac{\Delta\nu_{RF}}{\tau_0 [1 - D(\lambda_0)\lambda_0 c/n]}. \qquad (6)$$

The Sellmeier equation [28] based on Cauchy's equation for modeling dispersion [29] describes the dispersion of light. For the SMF-28 fiber, the chromatic dispersion is the differentiated Sellmeier equation

$$D(\lambda_0) \approx \frac{S_0}{4}\left[\lambda_0 - \frac{\lambda_{0D}^4}{\lambda_0^3}\right], \text{ps/(nm.km)}, 1200\,\text{nm} \leq \lambda_0 \leq 1600\,\text{nm}, \qquad (7)$$

where $S_0$ [0.092 ps/(nm$^2$.km)] is the zero dispersion slope, $\lambda_0$ (1302 nm< $\lambda_0$ <1322 nm) the zero dispersion wavelength, $\lambda_0$ (1542 nm) operation wavelength. Using Eq. 7 and differentiating Eq. 6, allow the determination of the non-linearity term [$d\nu'/(\nu' d\lambda_0)$ or $d\nu'/(\nu' d\nu_0)$] due to chromatic dispersion which is about $1.8 \times 10^{-5}$ (nm$^{-1}$) or $1.4 \times 10^{-7}$ (GHz$^{-1}$).

In practice, the residual fiber delay fluctuations due to temperature, vibrations and parasitic Fabry-Perot effects cause much stronger non-linearity than the chromatic dispersion. For instance, when we chirp the laser at a rate of 40.5 MHz/s stray reflections in the fiber system induce periodic changes of the optical frequency (Fig. 9a-1) with about 50 Hz peak-to-peak deviation for a 600 MHz tuning range. The procedure to determine $\tau_0$ is quite simple. We measure the frequency of the beat-note signal for a few seconds with the reference laser using the dead time free counter as shown before. We set sequentially $\Delta\nu_{RF}$ to 40 Hz and -40 Hz, and then we determine the chirp rates. This procedure cancels out the fiber delay time drift during the measurement. The timing of the measurement is very well known and only limited by the counter timebase. The difference of the chirp rates (32.44037 MHz) corresponding to an 80-Hz $\Delta\nu_{RF}$ is then calculated. Using Eq. 3 we precisely determine $\tau_0$ ($2.466064 \times 10^{-5}$ s). This accurate calibration holds for a short period of time (about 1 hour) due to temperature drift. On the long term an accuracy of $10^{-6}$ or better is achievable with temperature stabilization (in the level of 0.1 °C) and/or by using a fiber with lower thermal sensitivity. Then we chirp the laser at different rates ranging from 0.4 MHz/s to 40 MHz/s. The chirp rate error is the frequency difference between the measured and expected chirp rates. The expected rate equals to the sum of a constant laser drift measured prior to the chirp and a linear frequency ramp calculated using the $\tau_0$ and $\Delta\nu_{RF}$. The unavoidable frequency transient error (a few Hertz on millisecond time scale) due to finite response time of the control loop is well below the noise on the measurement system. We evaluate the relative chirp rate accuracy (Fig. 9b), which is defined by the ratio between the chirp rate error and the nominal chirp rate. Each measurement of the chirp rate lasts about 4 s, and the error bars are calculated by using the measured frequency instabilities of the chirped laser at 4 s integration time.

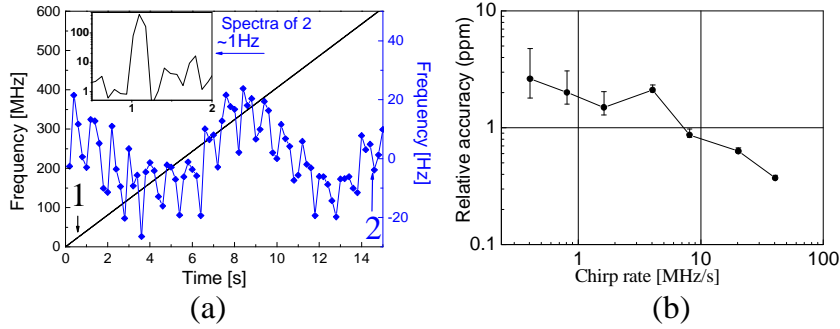

Fig. 9. (a-1) Beat-note signal frequency of the chirped laser vs. ULE-cavity stabilized laser, (a-2) residual frequency fluctuations (0.2 s gate time) after removing linear frequency fit (blue line+diamonds), (insert) the PSD of the residuals of a-2, (b) chirp rate accuracy with respect to the chirp rate.

This measurement demonstrates the potential of the technique, although the frequency drift (-2~2 kHz/s) still limits the accuracy especially at low chirp rates. However, with careful temperature control it is possible to reduce this effect at least below 100 Hz/s, or maybe even much less if temperature insensitive fibers become available. Beyond 10 MHz/s, the non-linear error is below 1 part per million (ppm). It is worth recalling that with the present drift rate of the laser, this result requires a frequency reference for periodically recalibrating the drift (each 5-10 minutes). The reference can be either a molecular transition [30] or a low drift cavity (< a few Hz/s).

The maximum chirp rate of our system is presently about 60 MHz/s limited by no optimized piezoelectric transducer response. We expect that well engineered set-ups could achieve at least one order of magnitude larger chirp rates still exhibiting low frequency noise.

*3.6 Applications of the agile laser*

Since the laser has low frequency noise and is digitally tunable, it can be frequency-locked or phase-locked with low control bandwidth. As a first example, it can be stabilized to an atomic transition or other stable references to achieve a low noise laser frequency reference exhibiting a good long-term stability. As a second example, it can be used as a low noise optical tracking oscillator for low orbit satellite (~1000-km) optical coherent ranging [31]. We also foresee this type of laser as a low cost clean-up optical tracking oscillator in long distance optical frequency distribution over fiber link [21,32,33].

**4. Conclusion**

We have developed an agile laser with ultra low frequency noise. We have obtained a frequency noise spectral density which competes with those obtained using ultra high finesse ULE Fabry-Perot cavities for Fourier frequencies higher than 20 Hz. In comparison with our previous non-chirped laser and at low Fourier frequencies we have improved the frequency noise by almost 10dB by careful thermal and mechanical isolation with probably large room for improvement.

Non-linear frequency error induced by the Fabry-Perot effect of stray reflections is approximately 50 Hz peak-to-peak in a 600-MHz tuning range and can definitely be reduced further. We have investigated the influence of RBS on the frequency noise performance of the chirped laser both theoretically and experimentally leading to a new insight on the fundamental limit of the frequency noise of the chirped laser.

**Acknowledgments**

We acknowledge funding support from the Agence Nationale de la Recherche (ANR BLAN06-3_144016) and Institut francilien de recherche sur les atomes froids (IFRAF). SYRTE is a Unité Mixte de Recherche (UMR 8630) of CNRS, Observatoire de Paris, and

## Appendix 1: Frequency noise PSD floor of the single RBS on a Michelson interferometer for a chirping laser

We assume here that the laser is swept at a constant rate $v'$, such than $v(t)=v(0)+v't$. At the distance $z$ from the input, elementary section of length $dz$ of the fiber induces a RBS with a coefficient $\rho(z)dz$. $\rho$ is a white random function [34], characterized by the autocorrelation function $<\rho^*(z_1)\rho(z_2)> = 2a_L a_B \delta(z_1-z_2)$, where $a_L$ (~3x10$^{-5}$/m) is the Rayleigh loss of the fiber per unit of length and $a_B$ (~1/500) is the backward coupling coefficient.

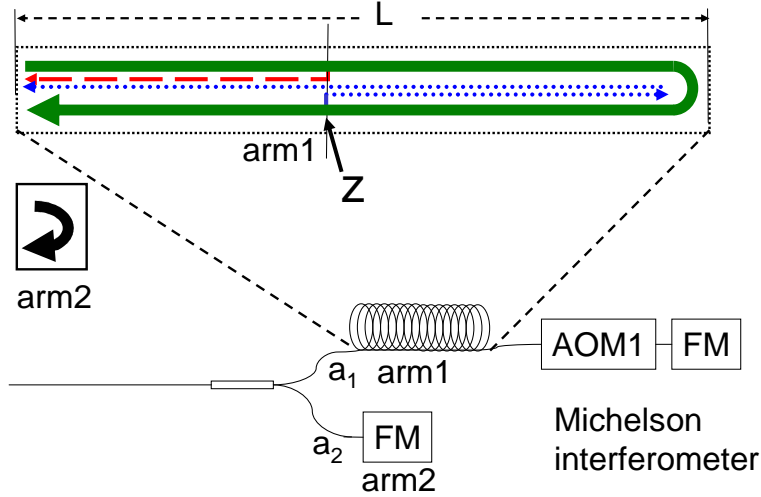

Fig. 1-1. The RBS effect on the laser frequency: AOM modulated reference signal (wide green line in arm1), short arm reference signal (wide black one), the RBS signals (red dash line, blue dot line); $a_1$ and $a_2$ are field amplitude coupling coefficients from the common input (in power).

Considering only RBS at $z$, four signals are combined at the output of the interferometer, $E_1(t)$, which is the signal coming from the reflection at the end of the arm #1 (long arm), $E_2(t)$, which is the signal coming from the reflection at the end of the arm #2 (short arm), $dE_{RBS1}(t,z)$ (red dash line) and $dE_{RBS2}(t,z)$ (blue dot line), which are the backscattered signal from $z$ of the forward signal and backward signal respectively. To simplify, the loss in the fiber is neglected. These signals can be expressed, in complex representation, as:

$$\begin{cases} \vec{E}_1(t) = a_1 a_{AOM1} E_0 \vec{e}_0 e^{i2\pi \int_0^{t-2nL_f/c}[v(t')+2f_{AOM1}]dt'} \\ \vec{E}_2(t) = a_2 E_0 \vec{e}_0 e^{i2\pi \int_0^{t} v(t')dt'} \\ \vec{dE}_{RBS1}(t,z) = a_1 \rho(z) E_0 \vec{e}_{RBS1}(z) dz\, e^{i2\pi \int_0^{t-2nz/c}[v(t')]dt'} \\ \vec{dE}_{RBS2}(t,z) = a_1 a_{AOM1}^2 \rho(z) E_0 \vec{e}_{RBS2}(z) dz\, e^{i2\pi \int_0^{t-2n(2L_f-z)/c}[v(t')+4f_{AOM1}]dt'} \end{cases}$$

(1-1)

where $a_{AOM1}$ is the double pass transmission coefficient of AOM1, $L_f$ the fiber geometric length, $n$ the fiber refractive index. $E_0$ is the input signal amplitude, $e_0$, $e_{RBS1}$ and $e_{RBS2}$ the polarization of the mirror directly reflected signals and the backscattered signals at the output. The beat-note signal between $E_1$ and $E_2$ is given by

$$I_{12}(t) = \eta a_1 a_2 a_{AOM1} E_0^2 e^{i(2\pi(2f_{AOM1} - \frac{2L_f n v'}{c})t}, \qquad (1\text{-}2)$$

where $\eta$ is a detection efficiency and the constant phase terms are ignored. The beat-note signals between $E_1$ and $dE_{RBS1}$ or $dE_{RBS2}$ are given by

$$\begin{cases} dI_{1RBS1}(t,z) = \dfrac{1}{\sqrt{2}} \eta a_1^2 a_{AOM1} \rho(z) dz E_0^2 e^{i(2\pi(2f_{AOM1} - \frac{2(L_f - z)nv'}{c})t} \\ dI_{1RBS2}(t,z) = \dfrac{1}{\sqrt{2}} \eta a_1^2 a_{AOM1}^3 \rho(z) dz E_0^2 e^{-i(2\pi(2f_{AOM1} - \frac{2(L_f - z)nv'}{c})t} \end{cases}, \qquad (1\text{-}3)$$

The factor $1/\sqrt{2}$ in the equation comes from the fact that $e_{RBS1}$, $e_{RBS2}$ and $e_0$ are random each other. Since $a_{AOM1}$ is generally less than 0.5, we neglect $dI_{1RBS2}$ in the following calculations. After mixing with the local signal at frequency $2f_{AOM1}-(2L_f n v')/c$, $I_{12}$ gives a large DC signal and $I_{1RBS1}$ give a small modulated signal at frequency $(2nzv')/c$. Consequently, the phase of the total error signals exhibit small modulations expressed as

$$d\theta_1(t,z) \approx \frac{1}{\sqrt{2}} \frac{a_1 \rho_1(z) dz}{a_2} e^{i2\pi \frac{2nzv'}{c} t}. \qquad (1\text{-}4)$$

To obtain the PSD of the total phase noise due to backscattering all along the fiber, we calculate the autocorrelation of the total phase noise by integration of the elementary phase noise contributions over $[0, L_f]$

$$\left\langle \iint_{[(0,0);(L_f, L_f)]} d\theta(t,z) d\theta^*(t',z') \right\rangle = \iint \frac{a_1^2}{2 a_2^2} \left\langle \rho(z) \rho^*(z') \right\rangle dz dz' e^{i2\pi \frac{2n}{c} v'(tz - t'z')}. \qquad (1\text{-}5)$$

Then by Fourier transform, we obtain the phase noise PSD:

$$\begin{cases} S_\theta(f) = \dfrac{a_1^2}{a_2^2} \dfrac{c}{8nv'} a_L a_B, \; f \leq \tau_0 v' = \Delta v_{RF} \\ S_\theta(f) = 0, \; f > \tau_0 v' \end{cases}, \qquad (1\text{-}6)$$

which is given from Eq. 1-5. The laser frequency noise PSD induced by backscattering

$$\begin{cases} S_v(f) = \dfrac{S_\theta(f)}{(2\pi \tau_0)^2} = \left(\dfrac{a_1}{a_2}\right)^2 \dfrac{a_L a_B c}{8n} \dfrac{1}{(2\pi \tau_0)^2 v'}, \; f \leq \tau_0 v' = \Delta v_{RF} \\ S_v(f) = 0, \; f > \tau_0 v' \end{cases}. \qquad (1\text{-}7)$$

**Appendix 2: Frequency noise PSD floor of the double RBS on a Michelson interferometer for a chirping laser**

Using the configuration shown in Fig. 7, we consider the influence of the second order Rayleigh backscattering that occurs at $z$ and $z'$ in the long arm. The expression of the three signals at the output of the interferometer is

$$\begin{cases} \vec{E}_1(t) = a_1 a_{AOM1} E_0 \vec{e}_0 e^{i2\pi \int_0^{t - 2nL_f / c}[v(t') + 2f_{AOM1}] dt'} \\ \vec{E}_2(t) = a_2 E_0 a_{AOM2} \vec{e}_0 e^{i2\pi \int_0^t [v(t') + 2f_{AOM2}] dt'} \\ \overline{d^2 E_{DRBS}}(t, z, z') = a_1 a_{AOM1} \rho(z) \rho(z') E_0 \vec{e}_{RBS}(z, z') dz dz' e^{i2\pi \int_0^{t - 2n[L_f + z - z'] / c}[v(t') + 2f_{AOM1}] dt'} \end{cases}.$$

(2-1)

The beat-note signal between $E_1$ and $E_2$ is given by

$$I_{12}(t) = \eta a_1 a_2 a_{AOM1} a_{AOM2} E_0^2 e^{i(2\pi(2[f_{AOM1} - f_{AOM2}] - \frac{2L_f n v'}{c})t)}, \quad (2-2)$$

and the beat-note signal between $E_2$ and $d^2 E_{DRBS}$ given by

$$d^2 I_{2DRBS}(t, z, z') = \frac{1}{\sqrt{2}} \eta a_1 a_2 a_{AOM1} a_{AOM2} \rho(z) \rho(z') dz dz' E_0^2 e^{i(2\pi(2[f_{AOM1} - f_{AOM2}] - \frac{2(L_f + z - z')n v'}{c})t)}.$$

(2-3)

This leads to the expression of the beat–note signal phase modulation

$$d^2 \theta(t, z, z') \approx \frac{1}{\sqrt{2}} \rho(z) \rho(z') dz dz' e^{i2\pi \frac{2n(z-z')v'}{c} t}. \quad (2-4)$$

After evaluation of the total autocorrelation function by integration over a triangle $(0,L,L)$ in the $z,z'$ plan, since $z'<z$, and Fourier transform, we obtain the frequency noise PSD as

$$\begin{cases} S_v(f) = \frac{S_\theta(f)}{(2\pi\tau_0)^2} = \frac{(a_L a_B)^2 c^2}{32\pi^2 n^2} \frac{1}{\tau_0 v'} \left(1 - \frac{f}{\tau_0 v'}\right), f \leq \tau_0 v' = \Delta v_{RF} \\ S_v(f) = 0, f > \tau_0 v' \end{cases} \quad (2-5)$$

The graphical representation of Eq. 1-7 and Eq. 2-5 is given in Fig 2-1.

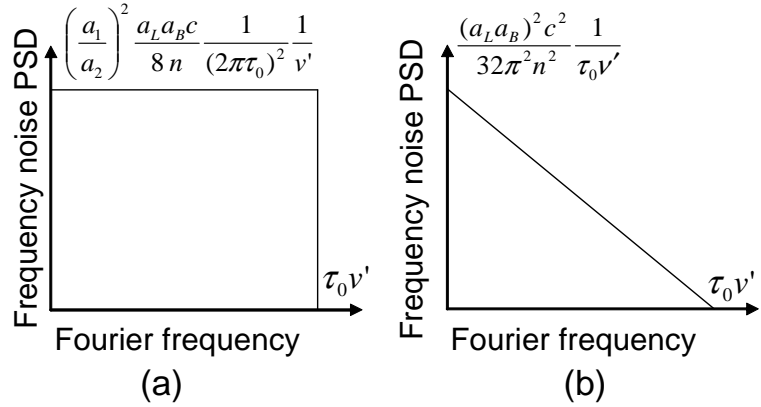

Fig. 2-1. The distribution of single RBS (a) and double RBS (b) induced frequency noise PSD (linear scales).